\documentclass[twocolumn]{aastex702}

\newcommand{\Msun}{\,{\rm M_\odot}}

\shorttitle{The Significance of Blue Companions to Little Red Dots}
\shortauthors{Pacucci \& Urry}
\accepted{\today}

\begin{document}

\title{More Numerous and Bluer: Companions of the Brightest Little Red Dots Differ from \\ Control Galaxies at $>3\sigma$, Hinting at Synchronized Black Hole Seed Formation}

\correspondingauthor{Fabio Pacucci} 

\author[orcid=0000-0001-9879-7780]{Fabio Pacucci}
\affiliation{Center for Astrophysics $\vert$ Harvard \& Smithsonian, 60 Garden St, Cambridge, MA 02138, USA}
\email[show]{fabio.pacucci@cfa.harvard.edu}

\author[0000-0002-0745-9792]{C. Megan Urry}
\affiliation{Department of Physics, Yale University, New Haven, CT, USA}
\email{meg.urry@yale.edu}

\begin{abstract}
The Little Red Dots (LRDs) are compact, red sources at $4 \lesssim z \lesssim 9$, likely powered by accreting massive black holes. Multiple studies have detected blue companions in their vicinity: we compare $253$ LRDs, uniformly selected across several JWST deep fields, against $>7000$ matched controls in the same mosaics. We find that $27.7\%$ of the brightest LRD quartile (in F444W) hosts a blue ($\beta<-1$) companion at $0.5-2$ kpc, versus $11.6\%$ of redshift-matched controls: a $\times 2.39$ excess, significant at $3.3\sigma$. Against controls additionally matched in brightness and compactness, the excess ratio is similar ($\times 2.17$), although constrained only at $1.4\sigma$ by the small size of that control pool. This phenomenon is limited to the brightest LRDs, as the full LRD sample pairs at a rate indistinguishable from controls. Moreover, the excess is confined to the close-pair window ($< 2$ kpc), outside which LRD environments are ordinary. Their companions are overwhelmingly ($92\%$) blue and significantly bluer than those of controls ($\beta\approx-2.32$ versus $-1.92$), rejecting a common parent distribution at $3.4\sigma$. The signal is thus directional in brightness, separation, and companion color, as expected for synchronized-pair seed formation. Therefore, we conclude that the association between the brightest LRDs and their close blue companions is physical, i.e., the two phenomena are causally connected: either one drives the other, or both emerge from the same synchronized formation event. A natural explanation is that the radiation from these young, star-forming neighbors enabled the direct collapse of the massive black holes powering the LRDs themselves.
\end{abstract}



\section{Introduction}
\label{sec:introduction}
Little Red Dots (LRDs) are a class of objects discovered by the \textit{James Webb Space Telescope} (JWST) during its first deep-field observations \citep{Kocevski_2023, Matthee_2023, Maiolino_2023_new}. The observational features of LRDs are peculiar.
First, they are very compact, typically unresolved, sources of light with effective radii $\lesssim 100$ pc \citep{Baggen_2023}.
Second, they are characterized by a V-shaped spectral energy distribution (SED) with increased emission at longer and shorter wavelengths \citep{Kokorev_2024_census, Kocevski_2024, Setton_2024}. These observational properties are used to effectively select LRDs \citep{Kocevski_2024}.

Originally interpreted as compact galaxies with high stellar densities \citep{Labbe_2023}, the paradigm began to shift when it was discovered that most LRDs are characterized by broad emission lines \citep{Greene_2023}, typically a signature of a massive black hole. These black holes appear extremely overmassive compared to the stellar mass of their hosts \citep{Pacucci_2023_JWST, Juodzbalis_2024_extreme, Jones_2025, Gupta_2026}, although many unknowns remain regarding the origin of their broad lines, leading to the possibility that their masses may be overestimated \citep{Rusakov_2025, Modelevsky_2026}. 

Currently, many interpretations for the nature of the LRDs rely on the presence of a massive black hole at their center \citep{Naidu_2025a, Pacucci_2026_DCBH}, although alternative explanations exist, such as involving globular clusters in formation \citep{Chisholm_2026}, supermassive stars with temperatures that can reproduce the LRD spectral features \citep{Nandal_2026}, or repeating tidal disruption events \citep{Bellovary_2025}.

An important piece of the puzzle in uncovering the nature of the LRD was recently unearthed with the discovery of abundant blue companions at projected separations of $0.5-5$ kpc \citep{Baggen_2026, Barger_2026}. A multiply-imaged LRD with an apparent blue companion was also discovered behind the galaxy cluster Abell 383 \citep{Golubchik_2026}.
These findings may be crucial: if the blue companions occur around LRDs much more often than in standard galaxies, there may be a causal connection between their presence and the LRD phenomenon. In other words, the LRDs may \textit{cause} the presence of the blue companions, or \textit{be caused} by it.

In particular, \cite{Baggen_2026} suggested that UV radiation from star formation, active in the blue companions, would enable the direct collapse of a massive black hole inside the nearby LRD. Direct collapse black holes (DCBHs, \citealt{Loeb_Rasio_1994, Oh_Haiman_2002, Bromm_Loeb_2003, Volonteri_2005, Begelman_2006, Lodato_Natarajan_2006}) were originally proposed as a fast formation pathway to explain the detection of $\gtrsim 10^9 \Msun$ supermassive black holes (SMBHs) at $z \gtrsim 6$ \citep{Fan_2001, Mortlock_2011, Wang_2021_quasar}, which would otherwise require challenging and sustained accretion rates, possibly at super-Eddington levels.
DCBHs, formed from the collapse of pristine (i.e., not metal-enriched), atomic-cooling \citep{Omukai_2001} halos, would have an initial mass in the range $\sim 10^4-10^6 \Msun$ \citep{Ferrara_2014}, significantly speeding up their growth process to reach the $\gtrsim 10^9 \Msun$ SMBH masses.

However, a condition sine qua non for the formation of a DCBH is the irradiation of the primordial, atomic-cooling halo by a significant flux of Lyman-Werner (LW) photons ($11.2-13.6$ eV, \citealt{Draine_1996}), which efficiently dissociates H$_2$ molecules via the Solomon process, hence preventing cooling and fragmentation. Three ingredients are thus necessary to form a DCBH: (i) pristine gas, (ii) an atomic-cooling halo, (iii) irradiation by a sufficient flux of LW photons. Although estimates of the minimum required LW flux vary widely \citep{Omukai_2001, Dijkstra_2008, Latif_2014_UV, Habouzit_2016, Pacucci_2017_CR7}, this condition is considered the most constraining for the collapse, as it would limit the locations where DCBHs can form. Recent simulations, however, indicate that substantially lower fluxes may suffice \citep{Hazlett_2026}.

Notably, \cite{Baggen_2026} found that the majority of the blue companions in their LRD sample already provide LW fluxes above the classical threshold; if the lower threshold of \cite{Hazlett_2026} is adopted, every companion in their sample would qualify.
The presence of these blue companions is reminiscent of the ``synchronized pair'' scenario of \cite{Dijkstra_2008, Visbal_2014_pair, Regan_2017}, in which a central halo and a sub-halo collapse nearly simultaneously, and the intense LW radiation from the first, star-forming halo triggers the formation of the DCBH in the companion halo.
Independently, \cite{Pacucci_2026_DCBH} recently suggested, using a series of radiation-hydrodynamic simulations, that the observational properties of LRDs can be entirely explained by identifying them with DCBHs.

However, normal galaxies tend to be clustered. Hence, the key question is: are the LRDs special? It is entirely possible that any object—LRD or ordinary galaxy at the same redshift—might have blue companions nearby. 
To confirm the exciting possibility that blue companions actually provide the necessary LW photons to trigger the DCBH formation, it is crucial to show that LRDs are significantly more likely to have blue companions nearby compared to a matched control sample of galaxies. 

This comparison must use an appropriate projected separation. The $0.5-5$ kpc range over which companions are typically reported (e.g., \citealt{Baggen_2026}) mixes two physically distinct regimes: a close-pair regime, where two halos are directly interacting (or even co-forming), and a wider environment, more connected to ordinary galaxy clustering. The distinction is crucial for the DCBH interpretation, because the LW flux received by the prospective collapsing halo at a distance $d$ scales as $d^{-2}$: only the closest separations enable the collapse.

Another important aspect deserving careful design is the choice of the control sample. The abundance of detected neighbors depends on many parameters, including redshift, brightness, and source compactness. Remarkably, these are properties in which the LRDs differ systematically from typical galaxies at the same cosmic epoch \citep{Kokorev_2024_census, Kocevski_2024, Taylor_2024}.

In this Letter, we compare a sample of $253$ LRDs to a carefully matched sample of galaxies, constructed as described in \S \ref{sec:data_methods}. Our results are described in \S \ref{sec:results} and discussed in \S \ref{sec:disc_concl}.
Throughout this Letter, we adopt a flat $\Lambda$CDM cosmology, with $H_0 = 70 \rm \, km \, s^{-1} \, Mpc^{-1}$ and $\Omega_m = 0.3$; across the redshift range of our sample, $1''$ corresponds to $4.6-7.0$ kpc of projected separation.

\section{Data and Methods}
\label{sec:data_methods}
We design a controlled experiment to compare the occurrence of blue close companions around LRDs versus normal galaxies. Specifically, we extract identical companion statistics around: (i) LRDs, (ii) Tier-1 control galaxies, matched in redshift, (iii) Tier-2 control galaxies, matched in redshift, brightness, and compactness, and (iv) random blank-sky positions. All measurements are performed on the same mosaics, with the same pipeline and identical parameters, so any difference between samples cannot arise from the measurement itself.
In the following, we describe the process in detail.

\subsection{The LRD Sample}
\label{subsec:LRDs}

Our LRD targets are derived from the photometric catalog in \cite{Kocevski_2024}, which applies uniform compactness and color selection criteria across several deep JWST surveys: we use its sources in CEERS \citep{Finkelstein_2025}, the JADES GOODS-S footprint \citep{JADES_2023}, NGDEEP \citep{Bagley_2024}, and the PRIMER COSMOS and UDS mosaics \citep{PRIMER_2021}.

We select sources in the redshift range $4.0 \leq z \leq 8.5$, within which the NIRCam short-wavelength bands sample the rest-frame UV continuum of the companions. This selection allows a uniform measurement of the UV slope, $\beta$, across the full sample. The parent catalog contains $341$ LRDs; we exclude the $23$ in the strongly lensed Abell~2744 field, where magnification distorts projected separations, and the redshift selection leaves $273$ objects. Of these, we remove $20$ (all in the GOODS-S region) for incomplete mosaic coverage, leaving a final sample of $253$ LRDs. Spectroscopic redshifts are available for $27$ objects, while we adopt the photometric redshifts from \cite{Kocevski_2024} for the remainder. From this LRD sample, we also define the brightest quartile: the 65 LRDs with F444W $\leq 25.20$. Throughout this Letter, ``bright'' refers to observed F444W magnitude. The F444W $\leq 25.20$ threshold is the upper-quartile edge of the sample's F444W distribution. Because the sample spans $4 \leq z \leq 8.5$, an F444W-selected bright quartile is skewed toward lower redshift. The comparison is nonetheless redshift-controlled, since every LRD is compared to its own redshift-matched controls, and we explicitly test the dependence on distance-corrected luminosity in \S\ref{subsec:results_excess}.
At the redshifts of our sample, F444W probes the rest-frame optical, $\sim 4700-8900$~\AA, which is the red side of the SED.

\subsection{Imaging Data and Control Samples}
\label{subsec:data}
All measurements for companion detection and photometry are performed on the public \textit{grizli} \citep{Brammer_2023} reductions of the JWST deep fields, using every short-wavelength filter available.

We construct two control tiers. In Tier-1, we draw up to $30$ control galaxies per LRD from the photometric catalog of the same field, matched in redshift within $|\Delta z|/(1+z) \leq 0.05$, and, for added safety, excluding any source within $2''$ of a detected LRD. The overlapping halves of the PRIMER mosaics duplicate $163$ controls, which we remove, leaving $7427$ unique galaxies in $253$ redshift-matched subsets. Tier-2 is defined by matching each LRD to the subset of controls satisfying a redshift, brightness, and compactness criterion. The redshift criterion in this case is $|\Delta z| \leq 0.50$, which is more permissive than Tier-1 by design. Most redshifts entering the match are photometric (\S \ref{subsec:LRDs}), with uncertainties that approach this tolerance at these redshifts; a tighter cut would thus select controls primarily on noise. 

The brightness criterion requires first placing controls and LRDs on the same photometric system.
Total F444W magnitudes for the controls are taken from the field catalogs (total fluxes where available, aperture fluxes otherwise) and converted to AB magnitudes. To place controls and LRDs on the same photometric system, we use the LRDs themselves, which appear in both catalogs: in each field, a positional cross-match yields the median magnitude difference between the two systems ($+0.15$ to $+0.28$ mag), which we subtract from the control magnitudes. The object-to-object scatter around this median (i.e., $0.10-0.15$ mag) is the residual calibration uncertainty of any individual comparison. We exclude the 322 control galaxies lacking a valid flux in their field catalog from Tier-2 matching, but keep them for Tier-1. For the LRDs, we adopt the F444W values directly from the \cite{Kocevski_2024} catalog, which defines the reference photometric system.

The brightness criterion is $|\Delta m_{444}| \leq 0.50$ mag, conservatively set at $3-5\times$ the residual cross-calibration scatter above, while still constraining controls to be within a factor of $\lesssim 1.6$ in flux. Statistical photometric errors at the flux levels of our sample are far smaller than the tolerance.
In addition, a control is defined as compact if its catalog half-light radius does not exceed $1.15$ times the $90$th percentile of the LRD radii measured in the same field, so that the threshold is insensitive to the absolute size calibration of each catalog. LRDs with fewer than three matched controls are excluded from this Tier-2 analysis, to avoid small-number statistics.

Matching both in brightness and in compactness is fundamental because spurious compact ``neighbors'' can be artificially manufactured from the host's light via two distinct mechanisms. The first mechanism relates to the algorithm we use: at fixed brightness, an extended light profile is far more likely to fragment into multiple segments during deblending than a point-like one. The second mechanism is astrophysical: our detection bands sample the rest-frame UV, where extended star-forming galaxies resolve into kpc-scale clumps \citep{Claeyssens_2023}; such clumps are compact and UV-blue, and thus satisfy the criteria (\S \ref{subsec:companion_measurement}) designed to validate real companions. Both effects scale with the host's brightness and extent and are minimal for point sources (\S \ref{subsec:validation}). Since LRDs are mostly unresolved, controls matched in brightness alone would be systematically more contaminated by fragments of themselves.

\subsection{Companion Measurement}
\label{subsec:companion_measurement}
We detect and measure companion properties with a pipeline that performs source detection directly on the short-wavelength imaging in cutouts centered on each target. We run detection on the signal-to-noise-weighted stack of the short-wavelength bands (F115W, F150W, and F200W, plus F090W where available) at a threshold of $2.5\sigma$ above the local background, with a deblending contrast of $10^{-3}$ so that close pairs are split wherever the data allow, following standard practice for JWST photometric catalogs \citep{Rieke_2023_JADES, Weibel_2024, Weaver_2024}.

Each companion detection is characterized by its UV slope, obtained from a least-squares fit:
\begin{equation}
    \log f_{\lambda} = \beta \log \lambda + {\rm const} \, ,
\end{equation}
to aperture photometry in every available short-wavelength band for which the central wavelength falls in the rest-frame range $0.13-0.30$ $\mu$m at the redshift of the target. We estimate background and noise locally in a $0\farcs75$ half-width box centered on the companion \citep{Weaver_2024}.

A candidate is considered a validated blue companion if it satisfies two requirements: (i) its aperture flux exceeds $2\times$ the aperture noise in at least two of the rest-UV bands entering the fit, and (ii) its UV slope is in the range $-3.5 \leq \beta < -1$. The upper bound is our main criterion for selecting blue companions. The lower bound is a data-quality control. Stellar populations cannot be arbitrarily blue: even zero-age, dust-free, metal-free ones can only reasonably reach $\beta \approx -3.2$ \citep{Schaerer_2003, Bouwens_2010}. Therefore, a measured $\beta < -3.5$ cannot correspond to a real spectrum and instead suggests noise-dominated photometry, which is typically caused by a faint flux fluctuation in one band tilting the fit \citep{Bouwens_2010}. This constraint rejects $556$ candidates with otherwise valid photometry ($\approx 11\%$ of the detections satisfying the aperture flux criterion within the $0.5-5$ kpc annulus); these rejections occur at statistically consistent rates around LRDs and controls.

If the candidate passes criterion (i) but has $\beta \geq -1$, its information is retained, and it is counted as a red companion; candidates below the blue bound are discarded.
For every target, we record the number of valid companions, $N$, and the binary indicator of hosting at least one. This latter indicator is our primary statistic for this project, answering the question: does this galaxy/LRD have at least one blue companion?

We consider two nested separation windows: the full $0.5-5$ kpc annulus and a close-pair window at $0.5-2$ kpc. The complementary annulus (i.e., $2-5$ kpc) serves as a larger-environment control window. The inner edge, in practice, is set by the short-wavelength angular resolution: companions can be counted only at separations $\geq 0\farcs15$, corresponding to $\approx 0.7$ kpc at $z = 8.5$. The smallest separation recorded anywhere in the sample is, indeed, $0.713$ kpc. All close-pair fractions are therefore lower bounds with respect to studies resolving smaller separations with, e.g., gravitational lensing \citep{Baggen_2026}.

\begin{figure*}[ht!]
\centering
\includegraphics[width=\textwidth]{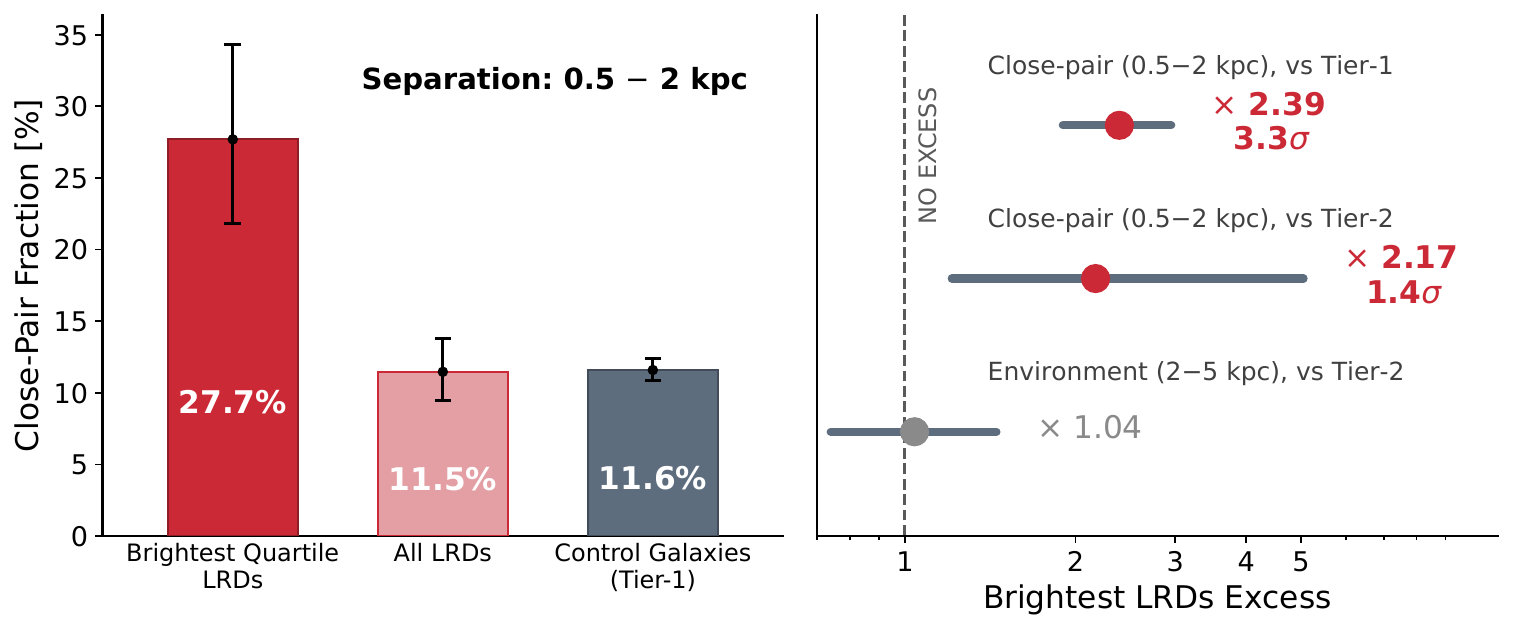}
\caption{Close-pair statistics at $0.5-2$ kpc. \textit{Left:} Fraction of sources hosting at least one validated blue ($\beta < -1$) companion at $0.5-2$ kpc, for the brightest LRD quartile (F444W $\leq 25.20$; $18/65$), all LRDs ($29/253$), and redshift-matched control galaxies (Tier-1). Error bars are the 68\% confidence interval. \textit{Right:} Excess of the brightest LRD quartile against the no-excess line, measured against redshift-matched controls (Tier-1, $\times 2.39$) and against controls additionally matched in brightness and compactness (Tier-2, $\times 2.17$). When measured against Tier-2 controls in the $2-5$ kpc environment, the LRDs show no excess ($\times 1.04$). The brightest-quartile excess for LRDs is detected at $3.3\sigma$ against coeval galaxies.}
\label{fig:excess}
\end{figure*}

Finally, each companion is cross-matched against the field catalog and its segmentation map and classified as: (i) \textit{substructure} when it falls within the same segment as the target (i.e., it is part of the target), (ii) \textit{independent} when it matches a distinct catalog source, or (iii) \textit{unmatched} when it has no catalog counterpart.
Our primary statistics retain all valid companions, including the ones classified as substructures. In fact, at the small separations of interest, a genuine close pair is frequently nearly merged into a single segment (particularly around bright compact hosts). Hence, discarding a same-segment detection would preferentially remove real pairs.

\subsection{Validation}
\label{subsec:validation}
Companion fractions include chance superpositions.
We quantify the chance superposition floor with a total of $300$ random positions per field, drawn uniformly within the catalog footprint and processed with the identical pipeline. A random sightline hosts a validated ``blue companion'' in $f_0 = 0.14-0.37$ of cases (depending on field depth; mean of $0.22$) over the full $0.5-5$ kpc annulus, and in $f_0 = 0.02-0.07$ (mean $0.04$) within the $0.5-2$ kpc close-pair window. This additive floor is shared by both LRDs and control galaxies. For this reason, we do not subtract it from ratios, as its presence can only bias them toward unity. Therefore, we conclude that the excesses we report are conservative: removing $f_0$ from the ratios would raise the bright-quartile close-pair excess from $\times 2.39$ to $\times 3$.

Two separate tests validate the measurements. First, the completeness test: we inject companion-like point sources around LRDs and matched controls, and recover them with the same pipeline. The fraction of injected sources that the pipeline re-detects is higher around LRDs than around controls, by $4-5$ percentage points at separations of $1-3$ kpc, with the two rates converging at $4$ kpc. This asymmetry is expected, as LRDs are faint in the short-wavelength detection bands, so an injected source near an LRD is superimposed on less host light than the same source near a control. Companions are thus found slightly more easily around LRDs. Correcting for this asymmetry, at first order, multiplies the close-pair ratio by $0.95$ (from $\times 2.39$ to $\times 2.27$): well within the quoted confidence interval. The completeness asymmetry, therefore, slightly amplifies the excess and cannot be its origin. 

Second, the artifacts test: we inject four flux-matched synthetic twins of each LRD at empty mosaic positions and re-run the entire pipeline. The twins are recovered in $966$ of $1012$ cases ($99\%$ in the bright quartile), and zero spurious close pairs are produced, bounding the spurious pair rate below $4.5\%$ per bright LRD ($1.2\%$ per LRD in the full sample) at $95\%$ confidence. In addition, as a safety measure, we visually inspect every system entering the close-pair statistics to confirm the results.

\section{Results}
\label{sec:results}

In this Section, we present the results of our study. First, we investigate whether LRDs are more likely than matched control galaxies to host a blue companion, and at which separations. Second, we study whether the companions they host differ in their properties from those of ordinary galaxies.

\begin{figure*}[ht!]
\centering
\includegraphics[width=\textwidth]{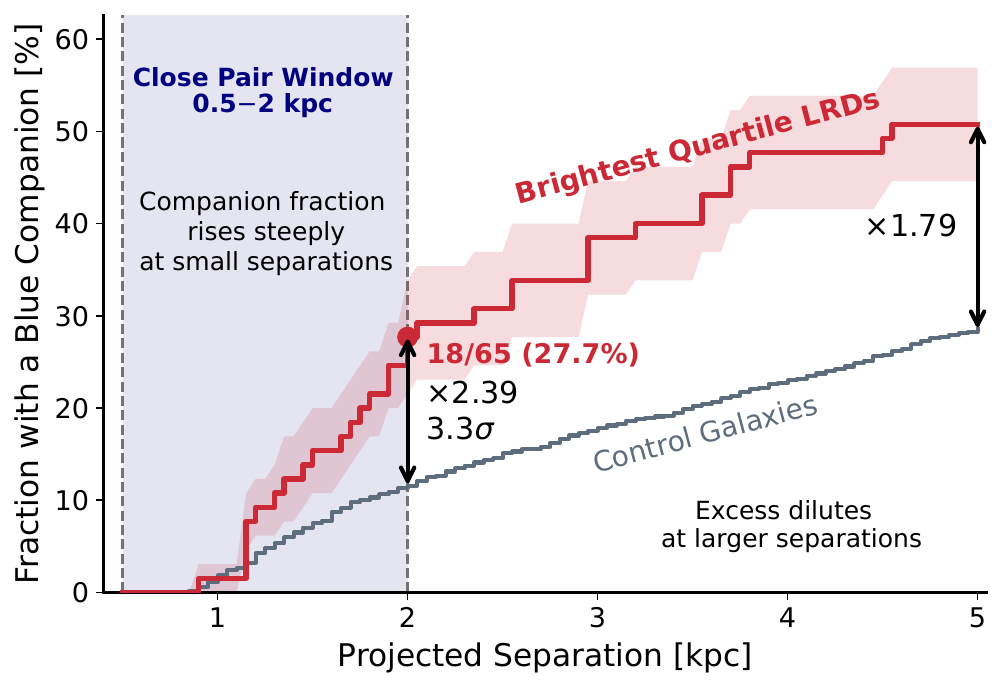}
\caption{Cumulative fraction of the brightest LRD quartile hosting a blue companion as a function of projected separation (red; shading is the $1\sigma$ confidence interval) and their Tier-1 control galaxies (gray). Both curves are empty below $\approx 0.7$ kpc: the $0\farcs15$ resolution floor corresponds to $\approx 0.7$ kpc at $z = 8.5$, comparable to the smallest separation recorded in the sample, $0.713$ kpc. The companion fraction rises steeply at $<2$ kpc, where $18/65$ ($27.7\%$) of the brightest quartile host a companion, a $\times 2.39$ excess over controls, significant at $3.3\sigma$. Beyond the close-pair window, the two populations accumulate companions at comparable rates, and the cumulative ratio declines to $\times 1.79$ by $5$ kpc. When measured in the $2-5$ kpc annulus against Tier-2 controls, the ratio is $\times 1.04$: at larger separations, the surroundings of the bright LRDs are indistinguishable from those of ordinary compact galaxies of the same brightness.}
\label{fig:separation}
\end{figure*}

\subsection{More Numerous Blue Companions, and Closer} 
\label{subsec:results_excess}
We begin our investigation with the close-pair window, $0.5-2$ kpc, whose statistics are summarized in Fig. \ref{fig:excess}.
Among the brightest quartile of our LRD sample ($65$ objects; hereafter, the bright LRDs), $18$ host at least one validated blue companion, while several host $\geq 2$. This finding leads to a blue companion pair fraction for bright LRDs of $27.7 \%$, with a $1\sigma$ confidence interval of $21.8-34.3\%$. The Tier-1 control galaxies (i.e., matched in redshift only) pair at $11.6\%$, when measured with an identical pipeline. The brightest LRDs are therefore ${\cal R}_z = 2.39$ times more likely to have a close-pair blue companion galaxy than redshift-matched field galaxies. This companion excess corresponds to a one-sided binomial probability of $p = 3.2 \times 10^{-4}$, or a $3.3\sigma$ excess.

The amplitude of the bright-LRD excess is also stable against the most restrictive tier of controls. In fact, against Tier-2 galaxies (i.e., matched in redshift, F444W brightness, and compactness), the ratio remains similar, at ${\cal R}_{z,m,c} = 2.17$. The two tiers play different roles. Tier-2 tests whether matching in brightness and compactness changes the ratio. We find that it does not: $\times 2.17$ against $\times 2.39$, consistent within the uncertainties. The Tier-2 constraint is only $1.4\sigma$ because few controls are both as bright and as compact as the LRDs. Importantly, the compactness criterion is what makes the comparison with equally bright galaxies fair, as LRDs are exceptionally bright for their small size \citep{Kokorev_2024_census, Kocevski_2024}.

This excess is a property of the brightest LRDs only. Across the full LRD sample, the pair fraction decreases to $29/253 = 11.5\%$, which is indistinguishable from the control rate (see Fig. \ref{fig:excess}). Hence, the physical connection between LRDs and their blue companions seems to operate in the brightest ones, not across the whole population. The dependence on brightness is continuous: the pair fraction declines monotonically across the four F444W quartiles ($27.7\%$, $11.3\%$, $4.8\%$, $1.6\%$), while the matched controls remain at $ 9-12\%$, a trend significant at $4.2\sigma$. The result is also unchanged if the quartile is defined in distance-corrected absolute magnitude ($\times 2.14$).

An instrumental origin is strongly disfavored (\S\ref{subsec:validation}): the synthetic-twin injections constrain the spurious pair rate to $<4.5\%$ per bright LRD, and the compact-control pair rate ($\approx 10\%$) bounds artifacts—both far below the observed $27.7\%$. 
The excess is carried by companions that share the host's catalog segment (\S \ref{subsec:companion_measurement}): $14$ of the $65$ bright LRDs host one, against $6.3\%$ of the controls (i.e., $\times 3.4$), whereas for companions cataloged as distinct sources the fractions are $9.2\%$ and $6.1\%$ (i.e., $\times 1.5$). However, the latter comparison has no discriminating power, because whether a source at $1-2$ kpc receives its own segment depends primarily on the host's brightness: $75-88\%$ of close blue companions of hosts brighter than F444W = 25.20 are same-segment, versus $13\%$ for hosts fainter than 28. The same-segment detections are separate peaks in the short-wavelength images at $0.9-2.0$ kpc, several PSF widths from the LRD, and none is reproduced by the synthetic twins (\S \ref{subsec:validation}).

The measured pair fraction also agrees quantitatively with comparable measurements in the literature, which find $\approx 21\%$ lensed LRDs with a companion in the separation range overlapping ours \citep{Barger_2026}. This value is consistent with our $27.7\%$.

In Fig. \ref{fig:separation}, we show how the dependence of pair fraction on separation localizes the phenomenon. In fact, the excess is established entirely within the innermost $2$ kpc. Within the close-pair window ($0.5-2$ kpc), the ratio already reaches its maximum value of ${\cal R}_z =  2.39$, while by $5$ kpc it has declined to ${\cal R}_z = 1.79$, as the bright LRDs and their controls accumulate blue companions at comparable rates.
The outer annulus (i.e., $2-5$ kpc) also confirms these conclusions: here, the bright-LRD ratio, measured against Tier-2 controls, is $1.04$. In other words, at larger separations, the surroundings of LRDs are indistinguishable from those of coeval compact galaxies of the same brightness.

The blue companions of the bright LRDs are thus: (i) more numerous, and (ii) systematically closer than those of control galaxies. Such a specific association, which is significant at $3.3\sigma$ against controls that share the same fields, depth, and measurement, is extremely unlikely to arise by chance.

\subsection{Overwhelmingly Blue, and Bluer, Companions} 
\label{subsec:results_colors}

\begin{figure*}[ht!]
\centering
\includegraphics[width=\textwidth]{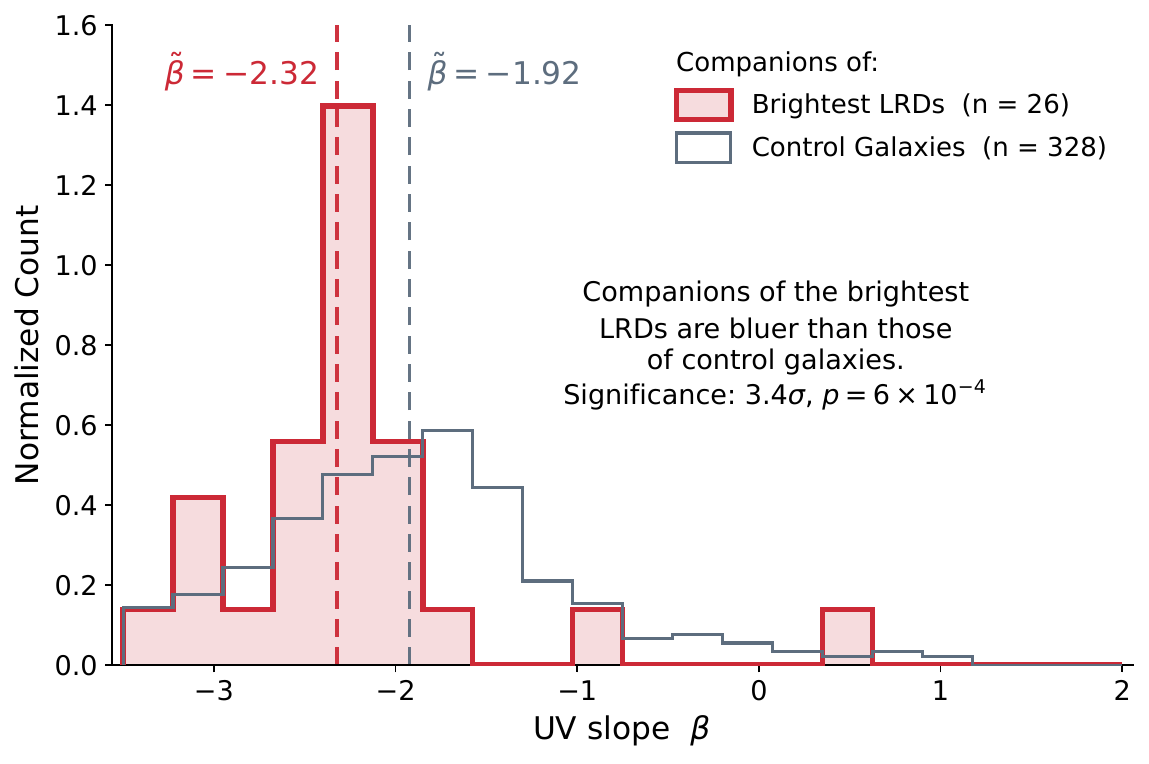}
\caption{Rest-UV slopes of all close companions with a measured $\beta$, with no color selection applied. Companions of the brightest LRDs are bluer than those of control galaxies, with median slopes $\tilde\beta = -2.32$ and $-1.92$, respectively (dashed lines). The difference is significant at $3.4\sigma$ ($p = 6 \times 10^{-4}$).}
\label{fig:colors}
\end{figure*}

Thus far, we have established that the brightest LRDs have more close blue companions. Now, we ask whether the bright LRDs and ordinary galaxies possess the same kind of companions. As we show below, the answer is a categorical no.

With no color selection, the brightest quartile of the LRD sample ($65$ objects) hosts $26$ close companions with a measured UV slope, compared with $328$ around their Tier-1 controls. Of these bright-LRD companions, $92\%$ (i.e., $24$) are UV-blue, with $\beta < -1$ (see Fig. \ref{fig:colors}). In addition, the LRD companions are characterized by a median slope of $\tilde{\beta} = -2.32$, while this value increases to $\tilde{\beta} = -1.92$ for those of the controls. A shift of $0.40$ in $\beta$ is remarkable: a two-sample Kolmogorov--Smirnov test rejects the null hypothesis that the two color distributions are drawn from the same parent population at $p = 6 \times 10^{-4}$: a $3.4\sigma$ significance. Hence, the companions of the bright LRDs are not only overwhelmingly blue, but also very significantly bluer than a random sample of companions of ordinary galaxies available at these redshifts.

The two companion samples are directly comparable: their detection-stack fluxes and the number of bands entering the fit are indistinguishable, and the typical uncertainty on an individual slope is $\sigma_\beta \approx 0.4$ for the LRD companions and $\sigma_\beta \approx 0.6$ for those of the controls. The median shift is thus comparable to a single-object uncertainty and reflects a sample property. The shift persists when the controls are restricted to the flux range of the LRD companions ($p = 1.4\times 10^{-3}$) and when only companions fitted in $\geq 3$ photometric bands are used ($\tilde{\beta} = -2.36$  versus $\tilde{\beta} = -1.93$, with $p = 3.8\times 10^{-4}$).

\begin{figure*}[ht!]
\centering
\includegraphics[width=\textwidth]{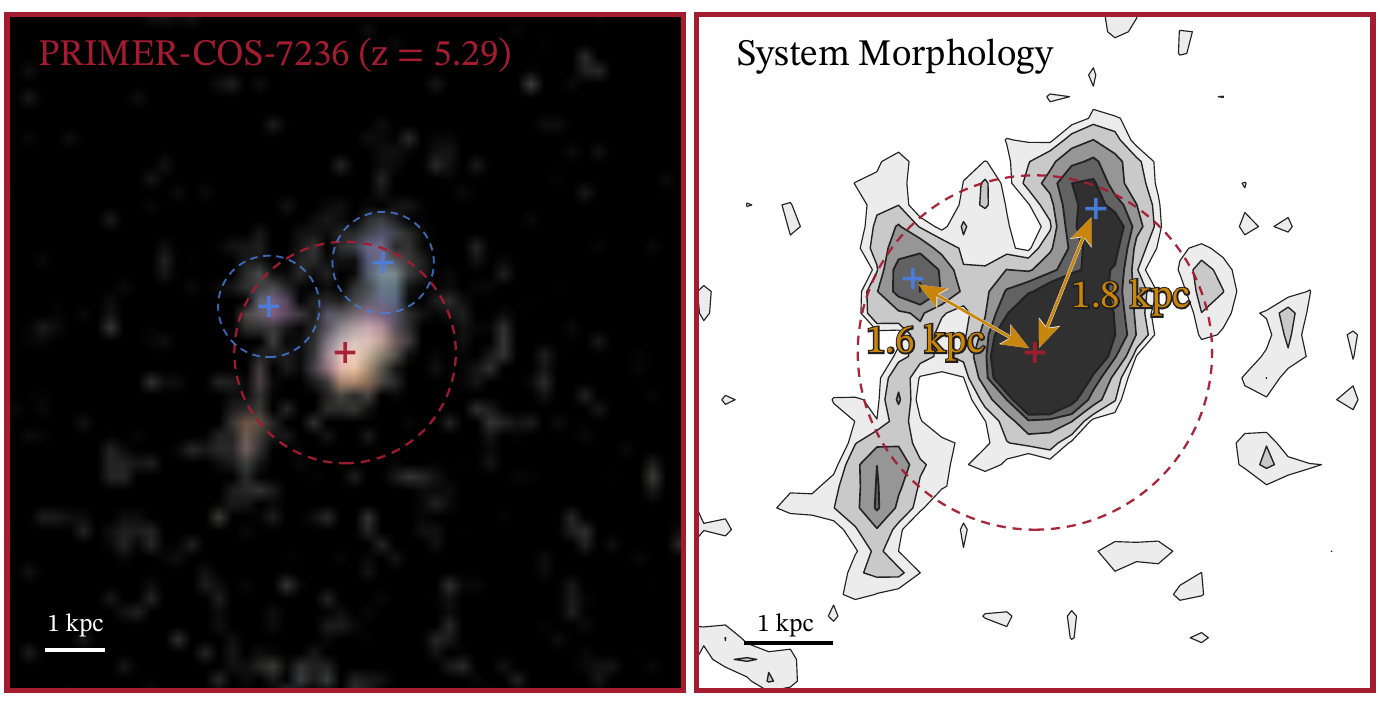}
\caption{A spectacular example of an LRD with two validated close blue companions: PRIMER-COS-7236 ($z = 5.29$). \textit{Left:} RGB composite image, with the red cross marking the LRD, the blue crosses marking the companions, and the dashed red circle marking the close-pair search radius. \textit{Right:} Detection map of the same system, with contours at 2, 3, 5, 8, and 12$\times$ the background RMS; a third validated blue companion (the structure to the lower left) lies just outside the window, at $\approx 2.3$ kpc. Such close companions are of the kind required to irradiate the central source with LW photons.}
\label{fig:pair}
\end{figure*}

The slopes of these LRD companions are those of young, low-extinction stellar populations: precisely the population capable of producing a significant flux of LW photons \citep{Baggen_2026}. An alternative origin for part of the blue light is discussed in \S \ref{sec:disc_concl}.
Figure \ref{fig:pair} displays a spectacular system: PRIMER-COS-7236 at $z = 5.29$. This LRD is flanked by two blue companions at projected separations of $\approx 1.8$ and $\approx 1.6$ kpc, both inside the close-pair window, as shown in the structural map.

It is thus remarkable that each constraint on the signal points to the same conclusion: (i) it appears only in the brightest LRDs, (ii) it is confined to separations below $2$ kpc, and (iii) it involves companions that are significantly bluer than those of ordinary galaxies. Clustering alone accounts for none of these restrictions. Each of them, however, is expected if a star-forming halo is irradiating the halo immediately next to it.

\section{Summary and Conclusions}
\label{sec:disc_concl}

This study was motivated by the discovery of abundant blue companions in the immediate vicinity of LRDs \citep{Baggen_2026, Barger_2026}; this finding may be key for understanding their nature. In fact, if star-forming companions occur near LRDs significantly more often than near ordinary galaxies, the UV radiation they provide may enable the direct collapse of massive black hole seeds, in the ``synchronized pair'' scenario \citep{Visbal_2014_pair}. Galaxies, however, are also clustered, and the mere presence of a companion is not conclusive. Claiming a causal connection between LRDs and blue companions requires a control.

We designed a controlled experiment, based on the photometric sample of LRDs in \cite{Kocevski_2024}, selecting $253$ LRDs at $4.0 \leq z \leq 8.5$ across several JWST deep fields. We measured their blue companion statistics against $>7000$ redshift-matched control galaxies, and against a more restrictive tier matched also in brightness and compactness. We performed every measurement on the same mosaics, with identical parameters. Our key findings are summarized here.

\begin{itemize} 
\item The brightest quartile of the LRD sample hosts a blue companion within $2$ kpc with a pair fraction of $27.7\%$. Identically measured, redshift-matched control galaxies pair at $11.6\%$. The excess is ${\cal R}_z = 2.39$, with a one-sided binomial probability of $p = 3.2 \times 10^{-4}$, significant at $3.3\sigma$.

\item The excess amplitude is essentially unchanged against the most restrictive control tier. Against controls matched in redshift, brightness, and compactness, the ratio is ${\cal R}_{z,m,c} = 2.17$, although at lower statistical significance ($1.4\sigma$) due to the smaller number of controls.

\item The phenomenon is restricted to the brightest LRDs. The full LRD sample has a pair fraction of $11.5\%$, indistinguishable from the control rate.

\item The phenomenon is equally restricted in separation. The pair excess declines from $\times 2.39$ within $2$ kpc to $\times 1.79$ by $5$ kpc. Additionally, the ratio declines to $\times 1.04$ in the $2-5$ kpc annulus, if measured against the most restrictive control tier. In other words, beyond the close-pair window, the surroundings of bright LRDs are those of ordinary compact galaxies of the same brightness.

\item The LRD companions are also remarkably different from companions of ordinary, coeval galaxies. Applying no color selection, the $26$ companions of the brightest LRDs have a median UV slope of $\tilde\beta = -2.32$, against $\tilde\beta = -1.92$ for the $328$ companions of the controls. A common parent distribution is rejected at $p = 6 \times 10^{-4}$, or $3.4\sigma$. Hence, the companions of the brightest LRDs are significantly bluer than those of the controls, and overwhelmingly blue in their own right: $24$ out of $26$ (i.e., $92\%$) have $\beta < -1$. This population of sources, being young and weakly obscured, can produce a substantial LW flux.

\end{itemize}

The emerging picture is surprisingly specific, as this Letter has shown that three independent restrictions (i.e., separation, companion color, and luminosity) all point in the same direction.
First, we have shown that the excess is confined entirely below $2$ kpc. Since the LW flux received by a candidate DCBH at distance $d$ scales as $d^{-2}$, only the closest companions can plausibly matter. 
Second, the companions we find are very significantly bluer (i.e., associated with young stellar populations) than those of ordinary galaxies at the same redshifts: exactly the population of sources maximizing the emission of LW dissociating radiation. 
Third, the excess appears at the bright end of the population, and nowhere else. This is expected if brightness (which tracks rest-frame luminosity for this sample, see \S \ref{subsec:results_excess}) traces youth. In fact, the brightest LRDs are likely the most recently activated systems, observed while the triggering starburst is still UV-bright and inside the close-pair window. Then, as these systems age, the signature is erased, because the companion fades or merges. The synchronized-pair scenario is thus the interpretation these trends support most directly, although these observations cannot establish the remaining conditions for direct collapse (i.e., pristine gas and an atomic-cooling halo).

This conclusion is entirely independent of \textit{assuming} a priori the direct collapse interpretation. Two distinct measurements, i.e., the occurrence rate of close companions and their color distribution, reject the null hypothesis at $>3\sigma$. In addition, this rejection occurs against very homogeneous control samples drawn from the same fields, observed to the same depth, and processed through the same pipeline. 

Based on this evidence, we conclude that the association between bright LRDs and their close blue companions must be physical. The two phenomena are \textit{causally connected}: either one drives the other, or both emerge from the same synchronized formation event. This is a constraint that models for LRDs must confront. A successful theory explaining their nature can no longer describe an isolated object, but it must also answer a fundamental question: why are the brightest of these sources found beside young, star-forming neighbors, at distances $\lesssim 2$ kpc?

We favor the synchronized-pair interpretation for two reasons. First, the companions resemble the blue clumps and satellites found around ordinary galaxies at these redshifts, only bluer, which points to star-forming systems rather than to a separate class of objects. Second, the scenario was not tuned to these data: the requirement that the LRD receive a strong LW flux from a young, star-forming halo predates our observations as a condition for direct collapse \citep{Dijkstra_2008, Visbal_2014_pair, Baggen_2026, Pacucci_2026_DCBH}, and it predicts companions that are close, blue, and found preferentially around the youngest, hence brightest, LRDs: these are the three trends we observe. The present data do not, however, determine the direction of the connection or the nature of the blue light. Part of it could be powered by the LRD itself, as young stars in an off-center clump or as low-density gas photoionized by the LRD, as proposed for the extended blue emission found around several LRDs \citep{Chen_2025b, Chen_2025}; the brightness and separation trends would follow in that case as well. Broad-band UV slopes cannot separate stellar from nebular continua; the equivalent width of [\ion{O}{3}]$+$H$\beta$, from medium-band photometry or spectroscopy, can. Under either reading, the association is physical.

Spectroscopy of the companions is the natural next step, as they are thus far identified photometrically. While the control comparison establishes the association statistically, only spectroscopy can confirm any individual pair and measure the companions' properties, such as the emitted LW flux.

To summarize, our study suggests that the close blue companions of bright LRDs are not incidental neighbors: the evidence collected clearly points to their being active participants in the phenomenon itself. With the bright LRDs, we may be witnessing, for the first time, the formation of massive black hole seeds caught alongside the very sources of light that made it possible.

\vspace*{1.2\baselineskip}

\begin{acknowledgments}
F.P. and C.M.U. thank the organizers and participants of the Sesto/Sexten Center for Astrophysics workshop on ``Deciphering the Growth of Supermassive Black Holes: Accretion, Mergers, and Cosmic Evolution'' for stimulating discussions and a productive research environment.
F.P. acknowledges discussions with Pierluigi Rinaldi and Dale Kocevski and support from NASA through Chandra Award No. GO3-24087A and JWST Award No. JWST-GO-03805.010-A.
C.M.U. acknowledges support from the National Science Foundation under grant No. AST-2407751.
Some of the data products presented herein were retrieved from the Dawn JWST Archive (DJA), an initiative of the Cosmic Dawn Center, funded under grant DNRF140.
\end{acknowledgments}

%

\software{\texttt{astropy} \citep{Astropy_2022},
          \texttt{numpy} \citep{Numpy_2020},
          \texttt{scipy} \citep{Scipy_2020},
          \texttt{matplotlib} \citep{Matplotlib_2007},
          \texttt{grizli} \citep{Brammer_2023}
          }



\bibliography{references}{}
\bibliographystyle{aasjournalv7.1}



\end{document}